\documentclass[twocolumn]{revtex4}
\usepackage{amsmath}
\usepackage{amssymb}
\usepackage{graphicx}
\usepackage{dcolumn}
\usepackage{natbib}
\usepackage{bm}
\usepackage{epstopdf}
\usepackage{multirow}
\begin{document}
\title{Landau level spectroscopy of surface states in the topological insulator Bi$_{0.91}$Sb$_{0.09}$ via magneto-optics}
\author{A. A. Schafgans}
\email{aschafgans@physics.ucsd.edu}
\affiliation{Department of Physics, University of California, San Diego, La Jolla, California 92093, USA}
\author{A. A. Taskin}
\affiliation{Institute of Scientific and Industrial Research, Osaka University, Osaka 567-0047, Japan}
\author{Yoichi Ando}
\affiliation{Institute of Scientific and Industrial Research, Osaka University, Osaka 567-0047, Japan}
\author{Xiao-Liang Qi}
\affiliation{Department of Physics, Stanford University, Stanford, California 94305, USA}
\author{B. C. Chapler}
\affiliation{Department of Physics, University of California, San Diego, La Jolla, California 92093, USA}
\author{K. W. Post}
\affiliation{Department of Physics, University of California, San Diego, La Jolla, California 92093, USA}
\author{D. N. Basov}
\affiliation{Department of Physics, University of California, San Diego, La Jolla, California 92093, USA}
\date{\today}

\begin{abstract}
We have performed broad-band zero-field and magneto-infrared spectroscopy of the three dimensional topological insulator Bi$_{0.91}$Sb$_{0.09}$. The zero-field results allow us to measure the value of the direct band gap between the conducting $L_a$ and valence $L_s$ bands. Under applied field in the Faraday geometry (\emph{k} $||$ \emph{H} $||$ C1), we measured the presence of a multitude of Landau level (LL) transitions, all with frequency dependence $\omega \propto \sqrt{H}$. We discuss the ramification of this observation for the surface and bulk properties of topological insulators.
\end{abstract}

\maketitle

Bi$_{1-x}$Sb$_{x}$ \cite{Galt-PR114-1396-1959,Heremans-PRB48-11329-1993,Lenoir-JPCS57-89-1996,Hsieh-1103.3413} was among the first materials predicted to be a three-dimensional (3D) topological insulator (TI) \cite{Kane-PRB76-045301-2007,Hasan-RMP82-3045-2010,Fu-PRL98-106803-2007,Teo-PRB78-045426-2008,Qi-PRB78-195424-2008}: a material with bulk insulating properties that supports conducting two-dimensional (2D) surface states. The appearance of nontrivial topological order is intimately tied to the band inversion that takes place as Bi is alloyed with Sb. Among the theoretical predictions now verified by experiment \cite{Hsieh-Nature452-970-2008,Hsieh-Science323-919-2009,Nishide-PRB81-041309R-2010,Roushan-Nature460-1106-2009} was for a free electron gas composed of spin-polarized quasiparticles existing at the 2D surface of the bulk 3D material, called surface states (SSs), formed when linear Dirac bands cross the Fermi energy ($E_f$). When a magnetic field is applied to the system, the SSs are thought to become gapped and spin-polarization reduced due to the time-reversal breaking field. In this work, best illustrated by the results in Fig. \ref{dRdH}, we demonstrate that all of the LLs observed in far-infrared magneto-optics obey Dirac-like dispersion as a function of applied field and are most likely due to optical transitions between LLs formed from the Dirac-like SS bands.

The sample in this study was a large ($\approx$ 1 cm$^{2}$) single crystal of Bi$_{0.91}$Sb$_{0.09}$ cut along the bisectrix [2\={1}\={1}] plane \cite{Taskin-PRB80-085303-2009}. This is the same plane (perpendicular to the [111] plane) in which quantum oscillations due to a 2D Fermi surface (FS) were observed in magneto-transport \cite{Taskin-PRB80-085303-2009,Taskin-PRB82-121302R-2010,Taskin-1009.4005}, implying the presence of topological SSs. We measured near-normal incidence reflectance in the far-infrared (30-700 cm$^{-1}$) as a function of temperature and applied field. In order to determine the zero-field optical constants, we measured reflectance between 30-8000 cm$^{-1}$ and variable angle spectroscopic ellipsometry between 4500 - 45000 cm$^{-1}$. We extracted the optical constants by performing a Kramers-Kronig constrained variational analysis using refFIT software, based on a multi-oscillator fit of the reflectivity data anchored by the dielectric function measured through ellipsometry \cite{Kuzmenko-RevSciInstrum76-083108-2005,Qazilbash-NatPhys5-647-2009}. These results were quantitatively similar to a full Kramers-Kronig inversion of the reflectivity data, which we show in the following figures.
\begin{figure}
\centering
\includegraphics[width=3.4in]{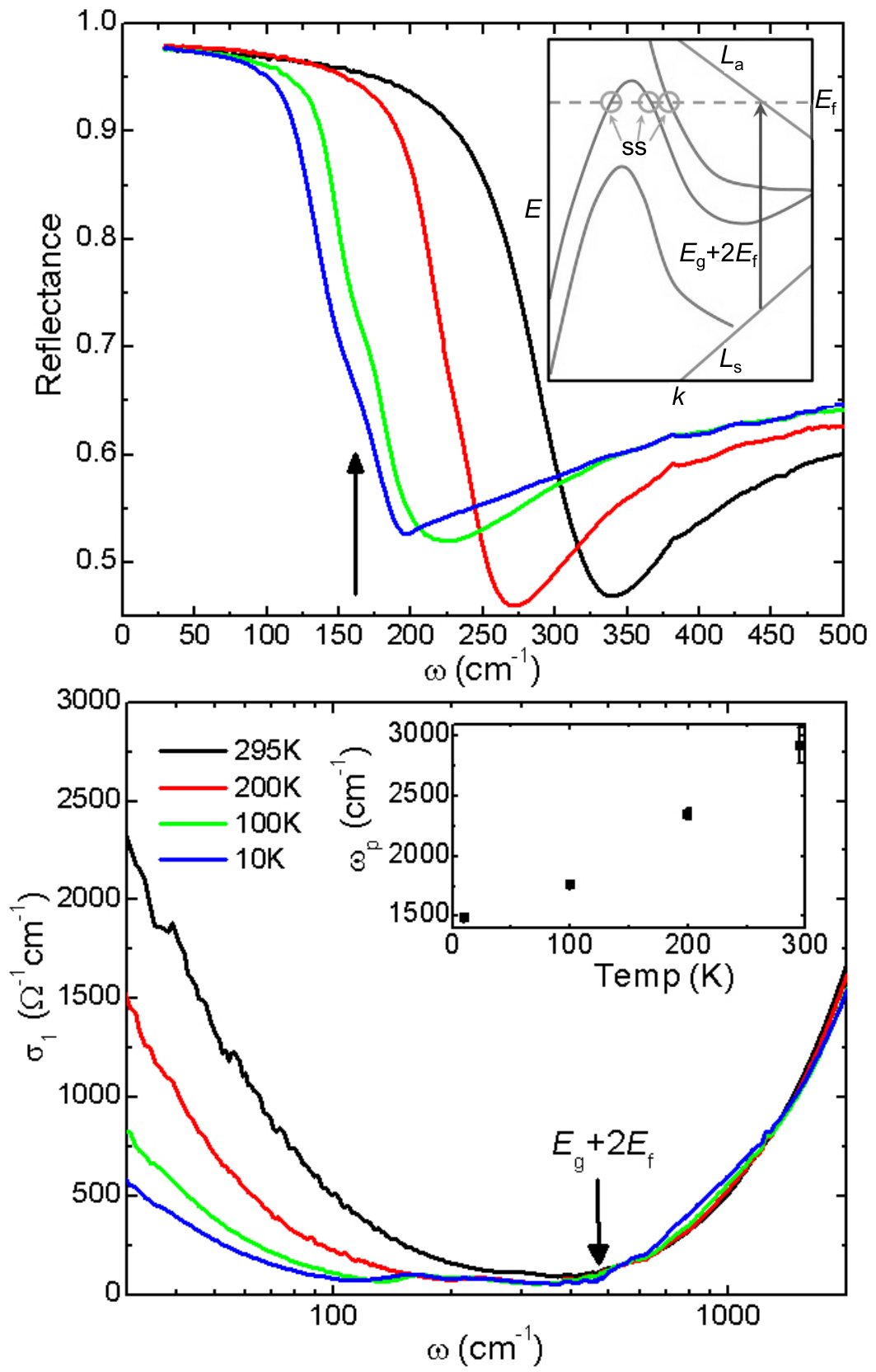}
\caption{\textbf{Top}. Reflectance vs. wavenumber in the far-infrared as a function of temperature, showing the Drude plasma frequency. The screened plasma frequency $\tilde{\omega_p}(10K)$ is shown with an arrow. \textbf{Top, inset} A cartoon showing the bulk $L$ bands, the Fermi energy $E_f$ defined from the bottom of the conduction band, the surface states formed when the surface Dirac bands cross $E_f$, and the necessary photon energy $E_g + 2E_f$ to create an interband transition. As far as we are aware, there have been no direct measurements of the SS bands along the C1 plane in Bi$_{1-x}$Sb$_{x}$. Transport measurements suggest there is at least one SS with Dirac dispersion, and in principle, there can be more (3, 5, ...). \textbf{Bottom}. The real part of the optical conductivity $\sigma_1(\omega)$ as a function of temperature, illustrating the decreasing Drude response at lowered temperature. The optical gap $E_g+2E_f$ is shown with an arrow, and the conductivity rapidly increases at frequencies above the optical gap. \textbf{Bottom, inset}. The Drude plasma frequency $\omega_p(T)$ obtained from modeling $\sigma_1(\omega)$ with the Drude-Lorentz model. The strong temperature dependence of $\omega_p(T)$ is likely the consequence of thermal activation.}
\label{Fig. 6.1}
\end{figure}

\section{Zero-field results}

The zero-field data are presented in Fig. \protect\ref{Fig. 6.1}, showing reflectance at four temperatures from room temperature to 10K (top panel). Unlike Bi$_{2}$Se$_{3}$ \protect\cite{LaForge-PRB81-125120-2010,Shushkov-PRB82-125110-2010}, the plasma edge in Bi$_{0.91}$Sb$_{0.09}$, formed by the free electron response and characterized by the plasma frequency $\omega_p$, demonstrates substantial temperature dependence. At low temperatures, the lineshape of the plasma edge contains structure that may indicate the presence of multiple plasma frequencies and is indicated with an arrow in Fig. \ref{Fig. 6.1}. With an Sb concentration of $x=0.09$, the alloy is a direct-gap ($E_g$) semiconductor between the bulk $L_a$ conduction band and the bulk $L_s$ valence band \protect\cite{Lenoir-JPCS57-89-1996}. This is evident in the real part of the optical conductivity $\sigma_1$ (Fig. \ref{Fig. 6.1}, bottom panel), where for photon energies greater than $\approx E_g+2E_f$, the onset of bulk interband transitions is observed: $E_g+2E_f$ = 470 $\pm$10 cm$^{-1}$, (indicated with an arrow in Fig. \ref{Fig. 6.1}, bottom panel). The upturn of the conductivity spectra at low frequencies is due to the metallic Drude response. As temperature is lowered, the Drude response is significantly reduced, indicating a vanishing bulk metallic Fermi surface ($\omega_p(T)$ is plotted in the inset). However, the remnant Drude response at the lowest temperature shows that the material retains bulk metallic properties.



\subsection{Location and size of Fermi Surface}

Taskin and Ando \cite{taskin} measured the volume of the bulk FS, formed by a set of three ellipsoids located at the \emph{L}-points of the Briullioin zone, to be $n= 8.1 \pm 0.2 * 10^{16}$cm$^{-3}$. The measured Hall coefficient implies a similar electron concentration of $n = 1.8 * 10^{17}$ cm$^{-3}$ \cite{taskin}. These data were obtained with samples from the same rod as the crystal used in this study and we therefore expect the properties to be quantitatively similar. The Drude model of the lowest temperature conductivity should produce comparable values between optics and transport only if the Fermi energy is located at an equivalent location in the bulk conduction band. This is because the plasma frequency of the Drude response is related to the number density of free carriers as: $\omega_p^2$ = 4 $\pi n e^2 / m^*$. Using the plasma frequency as measured via optics, $\omega_p$ = 1485 cm$^{-1}$, the resulting band mass is $m^*=0.0073m_e$. The plasma edge seen in optics is dominated by the shortest of the semimajor axes of the FS ellipoid, measured to be $k_f = 2.3 * 10^7$m$^{-1}$. The relationship $v_f = \hbar k_f / m^*$, where $v_f$ is the Fermi velocity, should provide the band velocity of the linear bulk conduction band along the fastest direction and therefore, the Fermi velocity in the fast direction of the bulk FS is $v_f = 3.65 * 10^5$ m/s.

Next, we invoke $E_f = \hbar k_f v_f$ in order to determine the separation between the bottom of the conduction band and the Fermi energy. This gives $E_f$ = 5.37meV = 42.5 cm$^{-1}$ above the bulk band gap, into the bulk conduction band. Based on the optical conductivity, since $E_g + 2E_f \approx 470 \pm$ 10 cm$^{-1}$, the magnitude of the bulk band gap between the $L_a$ and $L_s$ bands must be $E_g \approx 387 \pm$ 10 cm$^{-1}$. This is in excellent agreement with the data in Fig. 1 as well as previous photoemission studies. ($E_g$ is very sensitive to Sb content and above \emph{x}=0.04, $E_g$ grows with increasing Sb content \cite{Lenoir-JPCS57-89-1996}. In slightly higher doped Bi$_{0.9}$Sb$_{0.1}$, a lower bound of $E_g$=50meV=403cm$^{-1}$ was found using ARPES \cite{Hsieh-Nature452-970-2008}.) A cartoon of the bulk $L$ bands and the surface states is plotted in Fig. \ref{Fig. 6.1}, inset, showing the required incident photon energy to induce an interband transition between the bulk $L_a$ and $L_s$ bands. Also shown is a possible schematic of the SS bands \cite{Hsieh-Nature452-970-2008,Hsieh-Science323-919-2009,Zhang-PRB80-085307-2009}, illustrating that multiple SSs may still exist when the Fermi energy lies in the bulk conduction band. We explicitly note that most of the cited works have been performed or calculated for the [111] surface, whereas our present work is performed on the [2\={1}\={1}] surface. The surface state band structure along this surface has not presently been determined by photoemission studies.

\section{In-field reflectance data and modeling}

\begin{figure}
\centering
\includegraphics[width=3.4in]{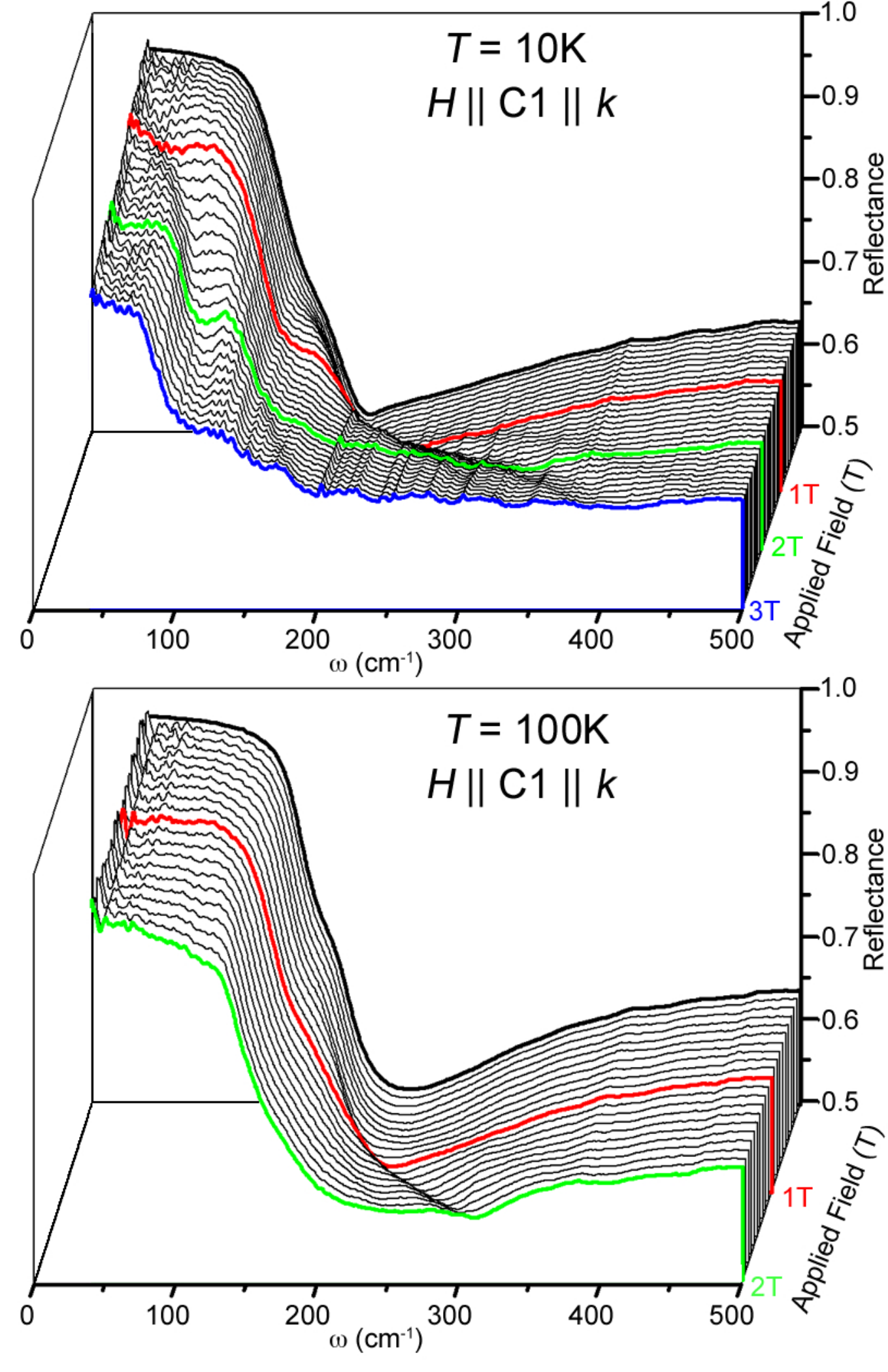}
\caption{Reflectance vs. wavenumber and applied field H $||$ C1 $||$ \emph{k}, at \emph{T} = 10K (top) and 100K (bottom). Each curve represents 0.1T increment in applied field, going from 0-3T at 10K and 0-2T at 100K.}
\label{Waterfalls}
\end{figure}

Turning to the in-field data, Fig. \ref{Waterfalls} shows reflectance in 0.1T increments at 10K (top) and 100K (bottom). The magnetic field was applied parallel to the bisectrix axis (C1), perpendicular to the bisectrix plane (C2-C3 plane), and parallel to the \emph{k}-vector of the incident electric field (Faraday geometry) in order to access the previously observed 2D FS deduced from quantum oscillation (QO) measurements \cite{Taskin-PRB80-085303-2009}. The large LL absorptions become visible above $\approx$ 0.3T, which likely correspond to bulk states, as well as other smaller field-dependent features. Models of the 10K data were used to extract values of the cyclotron resonance (CR) (see appendix). The models do not allow for the observation of the much weaker LLs due to the SS bands. Therefore, extracting LLs from the raw reflectance via modeling is not possible for anything other than the most prominent LLs because the surface state charge density is very small and therefore produces exceedingly weak features in reflectance.

\begin{figure}
\centering
\includegraphics[width=3.4in]{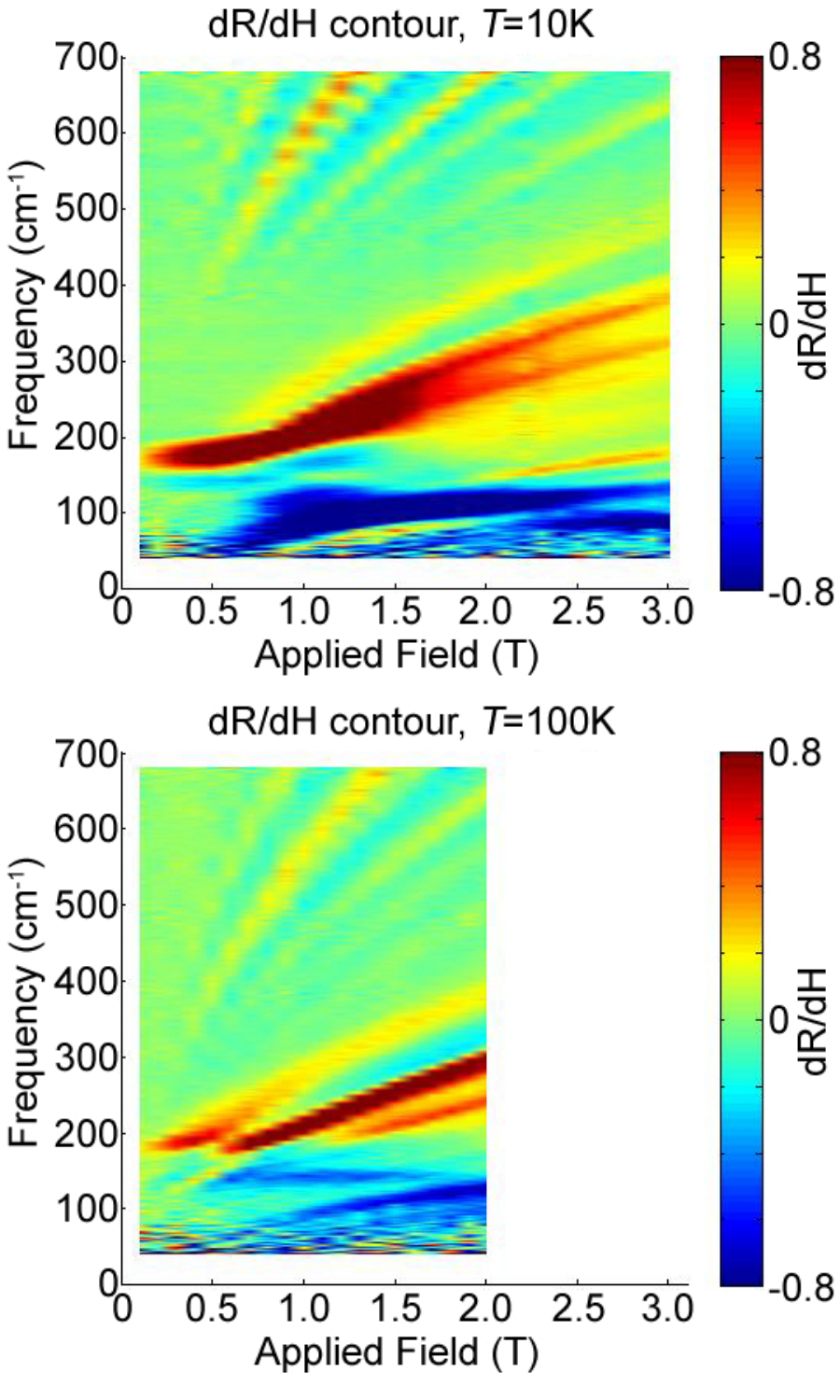}
\caption{Contour plots showing the derivative of the reflectance with respect to magnetic field, dR/dH, at \emph{T} = 10K (top) and 100K (bottom). LLs show up as peaks that disperse to higher wavenumber with applied field.}
\label{dRdH}
\end{figure}

Instead of modeling the data, differentiating the reflectance with respect to the applied field is a far more sensitive technique and allows access the subtle features corresponding to the 2D FS that are otherwise dominated by the bulk properties. Once we perform a derivative of the reflectance with respect to the applied field, dR/dH, the weakest LLs become quite evident. Indeed, these features are so small that they would have gone unobserved without utilizing this technique. Fig. \ref{dRdH} shows the dR/dH surface contour at at both $T$= 10K (top) and 100K (bottom), from $\omega$= 40-680 cm$^{-1}$ and $H$= 0-3T (at 10K) and $H$=0-2T (at 100K) with 0.1T increments. The value of dR/dH corresponds with the color scale. The intermediate field data are interpolated. As can be readily seen, there are multiple LLs, as well as features (dark red) that correspond to the field-dependent behavior of the bulk Drude plasma edge. All of the observed LLs display $\omega \propto \sqrt{H}$ dependence.

\section{Evolution of Landau levels in magnetic field}

A classical picture explaining the origin of the CR is that when a magnetic field is applied to a metal, free charge carriers become bound in orbits around the field lines, provided the mean-free path is sufficiently long. Such bound charge carriers create a resonance in the optical reflectance, dependent upon the strength of the applied field, the carrier's effective mass and number density. Since the band dispersion near $E_f$ is quadratic in most materials, the frequency of the CR and resultant LL transitions are $\omega_c \propto H$. As was noted early on, LLs in elemental bismuth have a non-linear field dependence due to the linear band dispersion of the $L_s$ and $L_a$ bands \cite{Wolff-JPhysChemSolids25-1057-1964,Brown-PRL5-243-1960,Vecchi-PRB14-298-1976,Heremans-PRB48-11329-1993}. Recently, the bulk $L_s$ valence band in Bi$_{0.9}$Sb$_{0.1}$ has been demonstrated to be linear using ARPES \cite{Hsieh-Nature452-970-2008}, in addition to the observation of linear Dirac bands that form SSs. 

In order to describe the energy spectrum of the LLs due to linearly dispersive bands, one must use a model derived from Dirac theory \cite{Wolff-JPhysChemSolids25-1057-1964,Furdyna-ProgPhys33-1193-1970}. For the bulk bands, the allowed energy states disperse as:
\begin{equation}
{E}={\pm(\frac{E_g^2}{4}+E_g(\frac{eH\hbar}{2m_c}(n+\frac{1}{2})\pm\frac{1}{2}g_0\mu_BH))^{1/2}}
\label{equation1}
\end{equation}
where $\mu_B$ is the Bohr magneton, $m_c$ is the experimental cyclotron mass, $n$ is the LL index of the bulk states, $g_0$ is the experimental \emph{g}-factor, and the first $\pm$ selects a LL in the conduction (+) or valence (-) band while the $\pm$ on the last term selects the spin state. This is different from the SSs in a three dimension topological insulator \cite{Liu-PRB82-045122-2010}, where the expression for the surface state LLs is:
\begin{equation}
\frac{E}{\hbar}=\frac{eHN}{m} \pm \sqrt{(-\frac{eH}{2m}+\frac{\mu_Bg_sH}{2\hbar})^2 + \frac{2eNHv_f^2}{\hbar}}
\label{equation2}
\end{equation}
\begin{equation}
\frac{E}{\hbar} = \frac{eH}{2m}-\frac{\mu_Bg_sH}{2\hbar}; N = 0
\label{equation3}
\end{equation}
Here, $g_s$ is the effective magnetic factor of the surface electrons, $m$ is a correction to the effective Drude mass, $N$ is the LL index of the SSs, and $\pm$ selects a LL above (+) or below (-) the Dirac point \cite{Liu-PRB82-045122-2010}. At low fields, the LLs disperse $\propto \sqrt{H}$, but as the field increases the linear terms will dominate.
\begin{figure}
\centering
\includegraphics[width=3.4in]{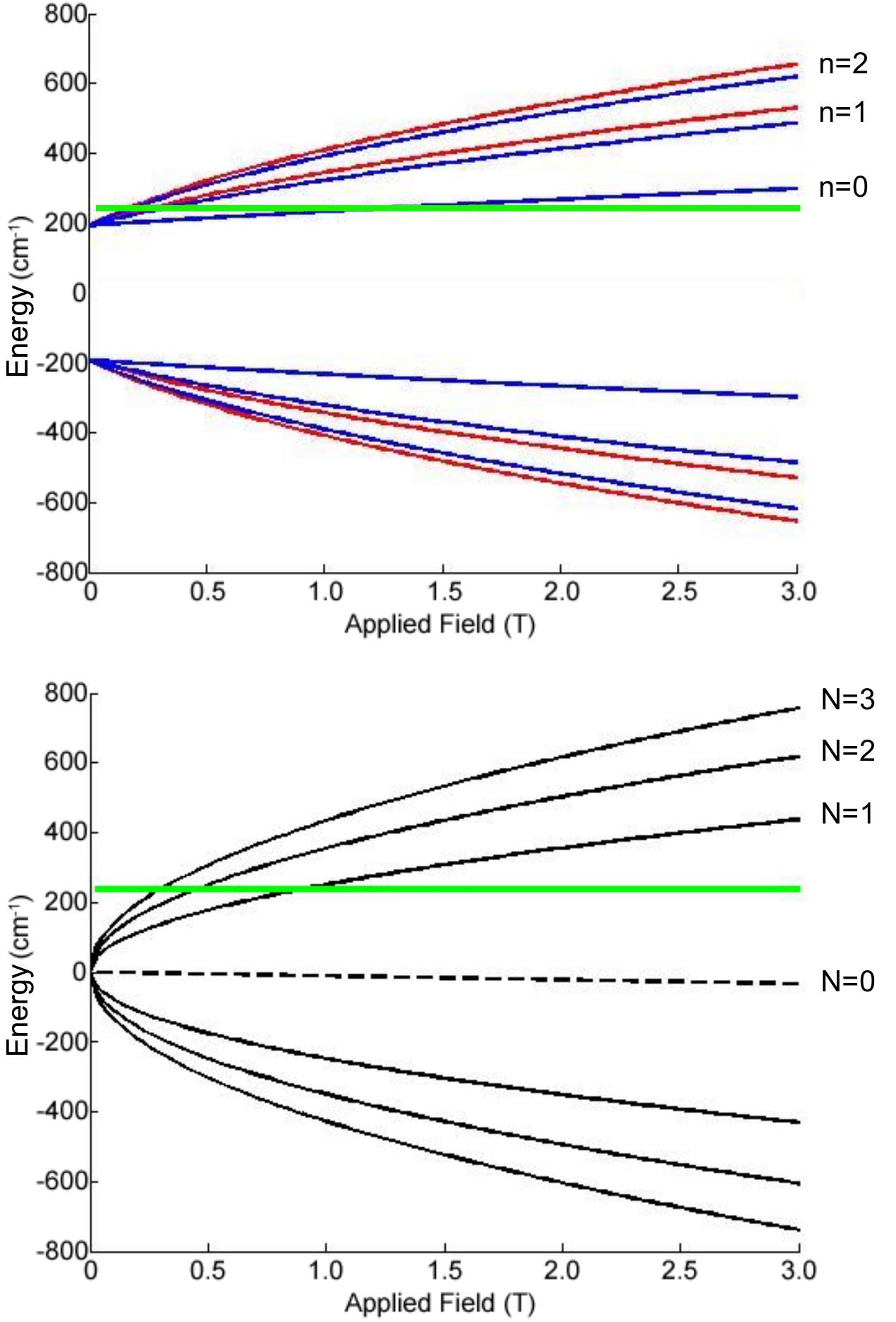}
\caption{Single particle LLs as a function of frequency for the bulk (top) and Dirac (bottom) bands, showing the n = 0-2 bulk LLs and the n = 0-3 Dirac LLs. The green horizontal line illustrates the location of the Fermi energy with respect to the single particle LLs ($E_f$= 45 cm$^{-1}$) as extracted from the zero-field data, as discussed above. The bulk LLs are spin-split, with the size of splitting determined primarily by the strength of spin-orbit coupling, which bears on the value of $g_0$. The red (blue) lines are spin up (down) LLs. Values used to determine the LLs are, for the bulk states: $m_c$=0.0073$m_e$, $E_g$=385 cm$^{-1}$, $g_0$=0.5/$m_c$. For the SSs, we used: $m$=1, $g_s$=80, $v_f$=8.5x10$^5$ m/s.}
\label{SPLLs}
\end{figure}

In Fig. \ref{SPLLs} we plot the single particle LLs for both the bulk (top) and SSs (bottom). There are several important features to be learned from these figures, as far as allowed optical LL transitions. First, as can be seen in the top panel, any interband LL transitions of the bulk states will extrapolate to a value $E_g$ at zero-field, while bulk intraband transitions will extrapolate to zero-frequency at zero-field. Second, bulk intraband transitions are not allowed once all of the bulk conduction LLs rise above $E_f$. (Using the parameters mentioned in the figure caption, this takes place near 1.5T. Of course, using a larger (smaller) effective mass means bulk intraband transitions are allowed to higher (lower) fields.) Third, all of the LL transitions due to the SSs (Fig. \ref{SPLLs}, bottom) will extrapolate to zero-frequency at zero-field. Fourth, the 0th LL is almost field independent for the parameters we use (see Fig. \ref{SPLLs} caption), and there is no spin-splitting. Fifth, LL transitions will not become allowed until the LLs cross $E_f$, and therefore the onset of optical transitions should be evident in the data. We note that the bulk band gap $E_g$ and Fermi level $E_f$ may not be field-independent, but any such field dependence should be a higher-order correction and will not significantly impact the results we show here. Also, the placement of the Dirac point, from where the surface state LLs disperse in field, may not be located in the center of the bulk band gap.

\begin{figure}
\centering
\includegraphics[width=3.4in]{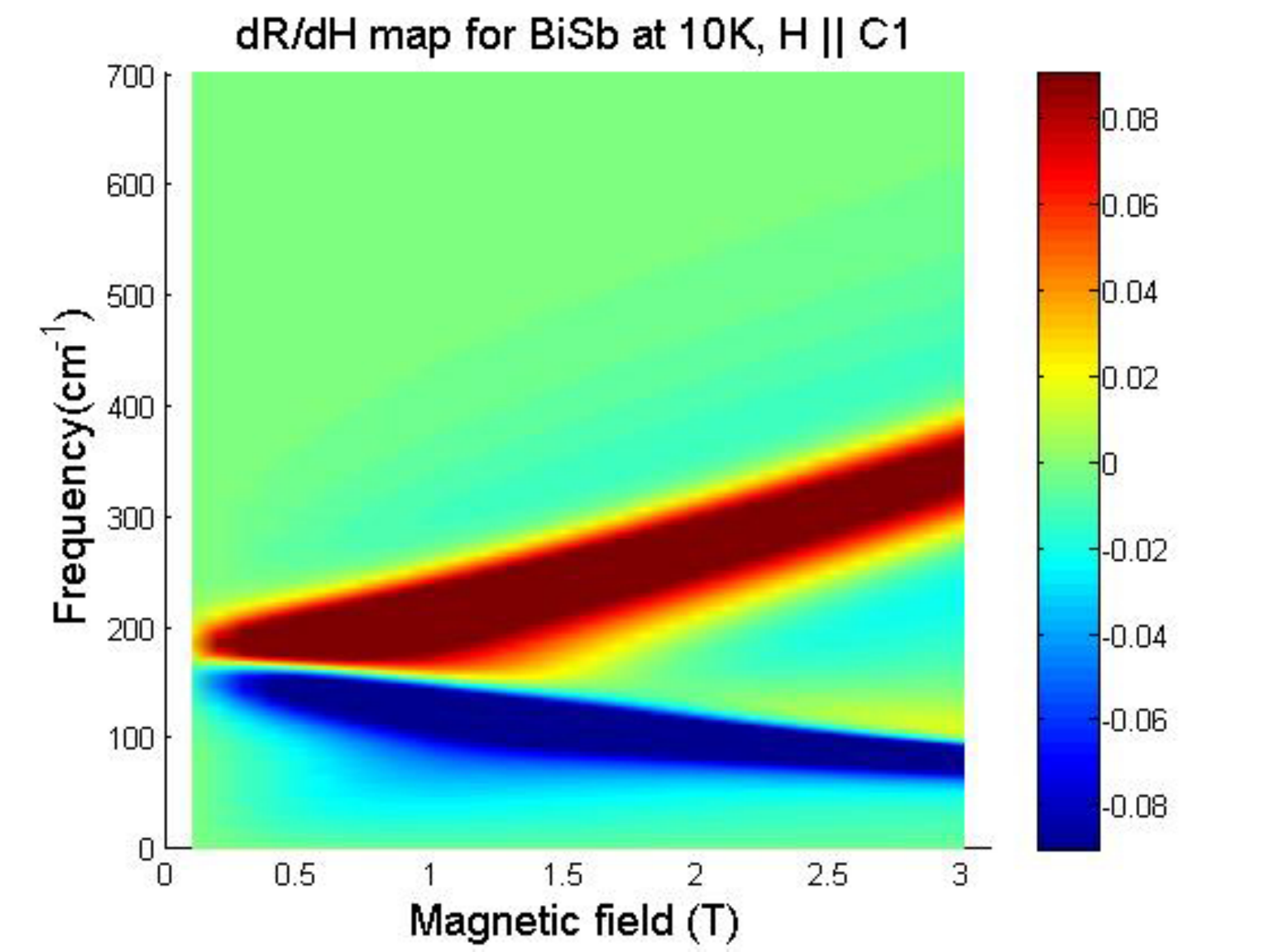}
\caption{The theoretical magnetic field dependence, plotted as a dR/dH contour, of a plasma edge of a metallic semi-conductor similar to what we observe in Bi$_{1-x}$Sb$_{x}$. The quantitative similarity with the data is suggestive.}
\label{PlasmaEdge}
\end{figure}

Additionally, we show that a simple model of the magnetic field dependence of the Drude plasma edge in metals \cite{Furdyna-ProgPhys33-1193-1970} can reproduce the behavior of the most prominent field-dependent features. Figure \ref{PlasmaEdge} demonstrates how the cyclotron active and inactive modes in magnetic field lead to a splitting of the plasma edge, where the cyclotron active mode moves the plasma edge higher in energy while the inactive mode suppresses the plasma edge. Qualitatively, such behavior is present in the data we show. However, more detailed analysis is needed in order to make definitive quantitative conclusions.



\section{Discussion}

Based on the information gained from the single particle LLs, we can make several immediate conclusions about the optical LL transitions in Bi$_{0.91}$Sb$_{0.09}$ presented in Fig. \ref{dRdH}. First, none of the observed LLs extrapolate to a zero-field value equal to the bulk band gap $E_g$. Therefore, based on eq. \ref{equation1}, none of the LLs are due to bulk interband transitions between LLs in the $L_a$ and $L_s$ bands. Second, as illustrated in Fig. \ref{SPLLs}, if there were any intraband LL transitions due to the bulk conduction bands present, these transitions should disappear by $\approx$ 1T. Therefore, none of the LLs can be understood to be due to the bulk bands within the paradigm presented by eq. \ref{equation1}. Third, the six highest energy LLs are observed to ''turn on" at finite field, consistent with the notion that once the LLs cross above $E_f$, transitions become allowed. 

\begin{figure}
\centering
\includegraphics[width=3.4in]{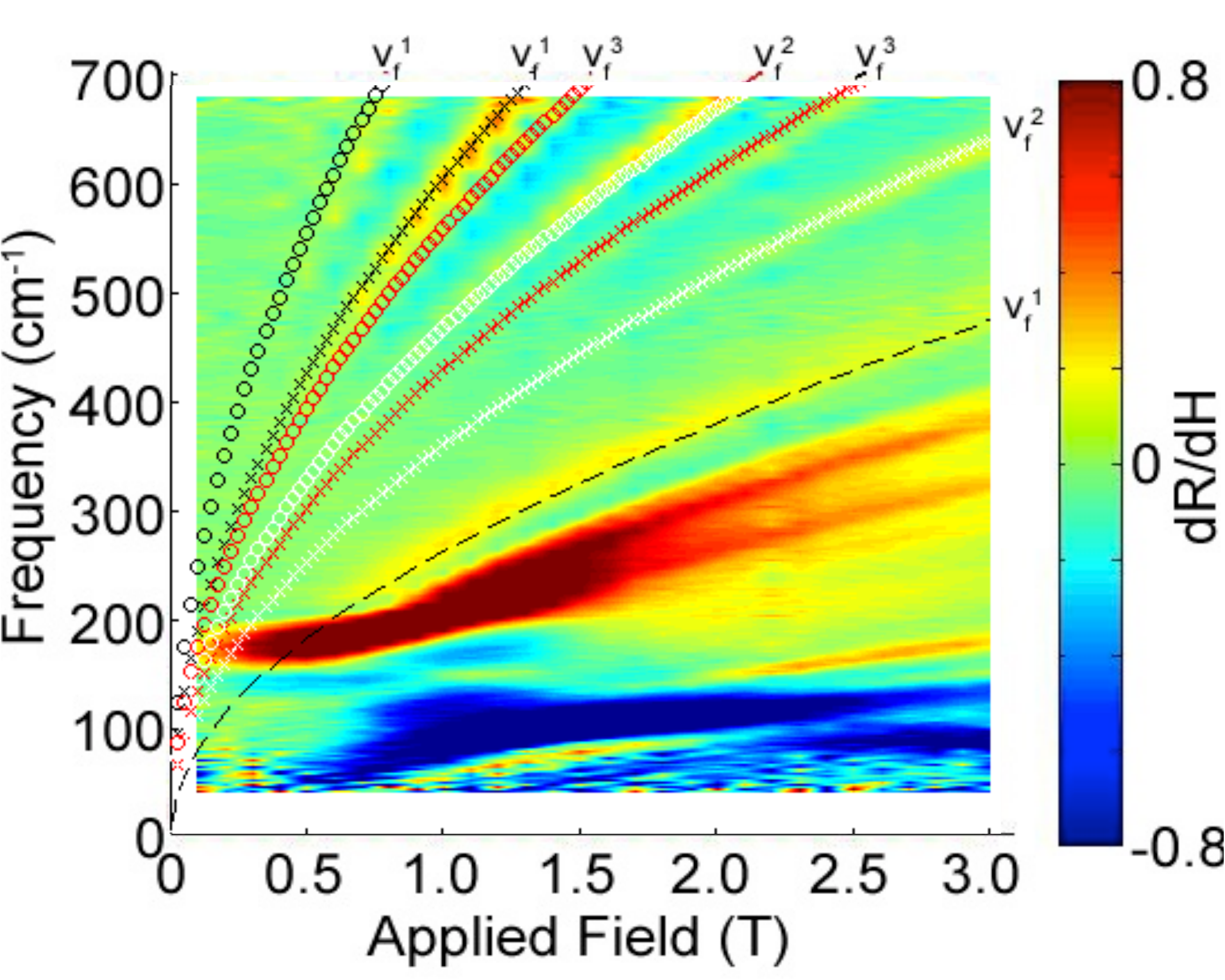}
\caption{Theoretical LL transitions of the form in eq.\ref{equation2} overlaid on the data at 10K. We have included three different Fermi velocities in order to explain all of the observed LLs. As labeled, $v_f^1$ = 8.5x10$^{5}$m/s (black symbols), $v_f^2$ = 6x10$^{5}$m/s (white symbols) and $v_f^3$ = 5x10$^{5}$m/s (red symbols) and the 0-1 (dashed line), 1-2:2-1 (crosses), and 2-3:3-2 (open circles) transitions are given. We did not correct for the Drude effective mass ($m$=1) and used a very moderate g-factor of the surface states ($g_s$=120).}
\label{SSLLs}
\end{figure}

A logical place to begin to quantitatively understand the LLs is to assume that the same surface state observed in quantum oscillation measurements on nearly identical samples is also contained in our data. Results from QO measurements determined the presence of a 2D FS with $v_f$ = 8.5x10$^{5}$m/s for very low fields ($<$1T) and at low temperature \cite{Taskin-PRB80-085303-2009}. In Fig. \ref{SSLLs} we show the predicted LL transitions due to a Dirac band with a Fermi velocity of $v_f^1$ = 8.5x10$^{5}$m/s for the n=0 to n=1 transitions (blue lines), 1-2:2-1 (black lines), and 2-3:3-2 (red lines) transitions. Therefore, we confirm previous measurements of the 2D FS.

Additionally, we find that by applying eq. \ref{equation2}, the number of LLs is greater than what can be explained with a single surface Dirac band. That is, in order to explain the data, we must include transitions that would exist in the presence of two additional Dirac bands with Fermi velocities determined by fitting the LLs: $v_f^2$ = 6x10$^{5}$m/s and $v_f^3$ = 5x10$^{5}$m/s. This presents one possible physical interpretation of the origin of the weakest LLs observed. In this scenario, it is interesting that the 0-1 transitions from $v_f^2$ and $v_f^3$ are either not present or completely overwhelmed by the behavior of the bulk plasma edge. ARPES studies of the [2\={1}\={1}] surface are required to determine the number of Dirac bands, and should be able to use our LL observations as a guide.

The frequency at which these transitions begin provides a clue as to the separation between the upper (+) and lower (-) LLs and ultimately provides a measure of the location of the Dirac point ($E_{Dirac}$) defined with respect to the bottom of the conduction band: the optical LL transitions of the SS bands will appear at $\omega$=$2(E_f-E_{Dirac})$, and will extrapolate to zero-frequency at zero-field. At \emph{T}=10K, the higher order transitions begin between 425-500 cm$^{-1}$ while the 0-1 LL transition is seen to begin at an energy equal to 1/2 of the higher order transitions: $(E_f-E_{Dirac})$. For instance, at $T=10K$, the 0-1 LL $v_f^1$ transition first appears near 225 cm$^{-1}$. For $E_f$ = 42.5 cm$^{-1}$, and therefore E$_{Dirac}$ is located $\approx$182.5 cm$^{-1}$ below the bottom of the bulk L$_a$ band, $\approx$ 11 cm$^{-1}$ above the center of the bulk band gap.

Interestingly, we do not observe bulk LL transitions across the gap \cite{Fuseya-JPSJ81-013704-2012}. This is somewhat surprising, given the sensitivity of our method for detecting LLs. One possibility is that the bulk transitions have short lifetimes, much shorter than the lifetimes of the SSs. The width of the Drude peak is one clue to the expected bulk LL lifetime, and roughly has a value of $\Gamma$ = 80-120 cm$^{-1}$, depending on modeling parameters. On the other hand, the lifetime of the SSs as determined from the width of the LL resonances is $\Gamma$ = 10-30 cm$^{-1}$, implying the SS carriers have lifetimes as much as 10 times longer than the bulk carriers. Transport measurements suggest a similar picture, where the mean free path of the bulk carriers was 16 nm while the SS carriers was 150 nm \cite{Taskin-PRB82-121302R-2010}. Our data, therefore, is consistent with the notion of scattering-protected SSs, while the bulk states are more strongly scattering.

In conclusion, we have detailed a magneto-infrared study of the topological insulator Bi$_{0.91}$Sb$_{0.09}$, demonstrating the presence of a multitude of Landau levels with $\omega \propto \sqrt{H}$ dependence \cite{Qi-PRB78-195424-2008}. After careful analysis, the LLs that we observe can only be understood to arise from Dirac-like bands within the bulk band gap. We therefore provide direct evidence for the existence of topological surface states in Bi$_{0.91}$Sb$_{0.09}$, coexisting with a bulk metallic Fermi surface. Future temperature and doping dependent LL studies as well as gated structures where the chemical potential can be tuned into various energy regimes of the band structure will help to further elucidate the origins of the observed phenomena we present.

We thank Gil Refael and Doron Bergman for enlightening discussions. Funding was provided by DARPA and FENA.

\section{Appendix}

\subsection{Reflectance models}

\begin{figure}
\centering
\includegraphics[width=3in]{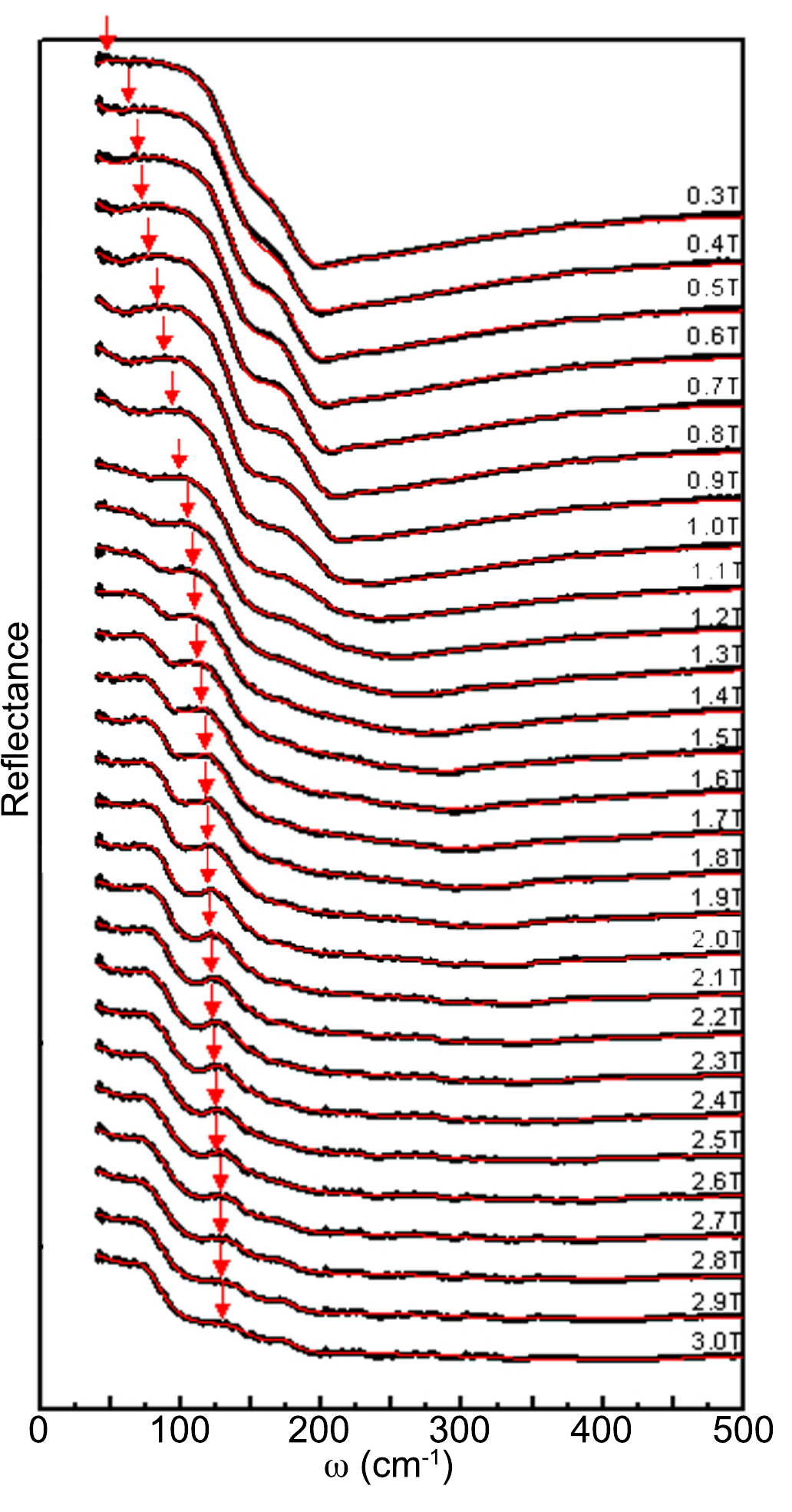}
\caption{Reflectance in 0.1T increments, overlayed with the corresponding model (red line) from RefFIT. }
\label{Models}
\end{figure}

\begin{figure}
\centering
\includegraphics[width=3.4in]{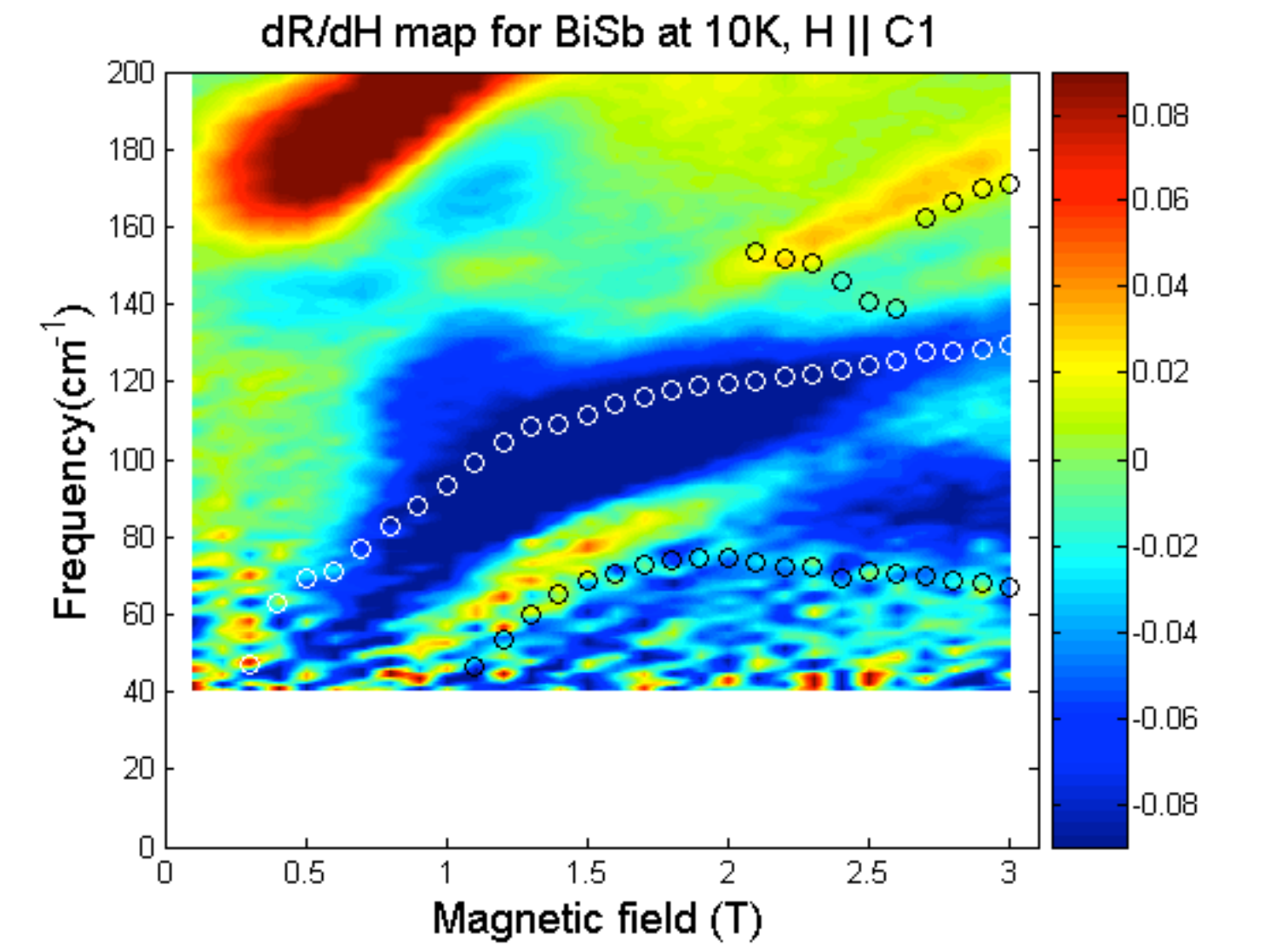}
\caption{dR/dH contour illustrating the three CR modes extracted from fitting the field dependent reflectivity. The white circles correspond to the CR mode that is shown with an arrow in Fig. \ref{Models}, while the black circles show two additional modes converged upon by the fitting routine.}
\label{dRdHModels}
\end{figure}

In order to extract the cyclotron resonance due to LL transitions, we attempted to model the far-infrared reflectance using refFIT \cite{Kuzmenko-RevSciInstrum76-083108-2005}. The resultant models are shown overlaid on the corresponding field-dependent reflectance data in Fig. \ref{Models}, where we point out the value of the modeling result for the primary cyclotron resonance with an arrow. The models do not allow for the observation the much smaller LLs due to the SS bands. At fields closer to 3T, fitting converged on the presence of three CR modes, however they did not display an intelligible systematic behavior. This may be due to the difficulty of fitting the raw reflectance and as such, we have not drawn conclusions based on these models. In Fig. \ref{dRdHModels}, we plot the values of the CR as determined by modeling the raw reflectance (circles), overlaid on the dR/dH contour.

\begin{thebibliography} {99}
\expandafter\ifx\csname natexlab\endcsname\relax\def\natexlab#1{#1}\fi
\expandafter\ifx\csname bibnamefont\endcsname\relax
  \def\bibnamefont#1{#1}\fi
\expandafter\ifx\csname bibfnamefont\endcsname\relax
  \def\bibfnamefont#1{#1}\fi
\expandafter\ifx\csname citenamefont\endcsname\relax
  \def\citenamefont#1{#1}\fi

\bibitem{Galt-PR114-1396-1959} J. K. Galt, W. A. Yager, F. R. Merritt, B. B. Cetlin, and A. D. Brailsford, Phys. Rev. \textbf{114}, 1396 (1959).

\bibitem{Heremans-PRB48-11329-1993} J. Heremans, D. L. Partin, C. M. Thrush, G. Karczewski, M. S. Richardson, and J. K. Furdyna, Phys. Rev. B \textbf{48}, 11329 (1993).

\bibitem{Lenoir-JPCS57-89-1996} B. Lenoir, M. Cassart, J.-P. Michenaud, H. Scherrer, and S. Scherrer, J. Phys. Chem. Solids \textbf{57}, 89 (1996).

\bibitem{Hsieh-1103.3413} D. Hsieh, Y. Xia, L. Wray, D. Qian, J. H. Dil, F. Meier, L. Patthey, J. Osterwalder, G. Bihlmayer, Y. S. Hor, R. J. Cava, and M. Z. Hasan, Cond. Mat. arXiv:1103.3413

\bibitem{Kane-PRB76-045301-2007} L. Fu, and C. L. Kane, Phys. Rev. B \textbf{76}, 045302 (2007).

\bibitem{Hasan-RMP82-3045-2010} M. Z. Hasan and C. L. Kane, Rev. Mod. Phys. \textbf{82}, 3045 (2010).

\bibitem{Fu-PRL98-106803-2007} Liang Fu, C. L. Kane, and E. J. Mele, Phys. Rev. Lett. \textbf{98}, 106803 (2007).

\bibitem{Teo-PRB78-045426-2008} Jeffrey C. Y. Teo, Liang Fu, and C. L. Kane, Phys. Rev. B. \textbf{78}, 045426 (2008).

\bibitem{Qi-PRB78-195424-2008} Xiao-Liang Qi, T. L. Hughes, and Shou-Cheng Zhang, Phys. Rev. B \textbf{78}, 195424 (2008).

\bibitem{Hsieh-Nature452-970-2008} D. Hsieh, D. Qian, L. Wray, Y. Xia, Y. S. Hor, R. J. Cava, and M. Z. Hasan, Nature \textbf{452}, 970 (2008).

\bibitem{Hsieh-Science323-919-2009} D. Hsieh, Y. Xia, L. Wray, D. Qian, A. Pal, J. H. Dil, J. Osterwalder, F. Meier, G. Bihlmayer, C. L. Kane, Y. S. Hor, R. J. Cava, and M. Z. Hasan, Science \textbf{323}, 919 (2009).

\bibitem{Nishide-PRB81-041309R-2010} A. Nishide, A. A. Taskin, Y. Takeichi, T. Okuda, A. Kakizaki, T. Hirahara, K. Nakatsuji, F. Komori, Yoichi Ando, and I. Matsuda, Phys. Rev. B \textbf{81}, 041309(R) (2010).

\bibitem{Roushan-Nature460-1106-2009} P. Roushan, J. Seo, C. V. Parker, Y. S. Hor, D. Hsieh, D. Qian, A. Richardella, M. Z. Hasan, R. J. Cava, and A. Yazdani, Nature \textbf{460}, 1106 (2009).



\bibitem{Taskin-PRB80-085303-2009} A. A. Taskin and Yoichi Ando, Phys. Rev. B \textbf{80}, 085303 (2009).


\bibitem{Taskin-PRB82-121302R-2010} A. A. Taskin, Kouji Segawa, and Yoichi Ando, Phys. Rev. B \textbf{82}, 121302(R) (2010).

\bibitem{Taskin-1009.4005} A. A. Taskin, Kouji Segawa, and Yoichi Ando, arXiv:1009.4005

\bibitem{Kuzmenko-RevSciInstrum76-083108-2005} A.B. Kuzmenko, Rev. Sci. Instrum. \textbf{76}, 093108 (2005).

\bibitem{Qazilbash-NatPhys5-647-2009} M. M. Qazilbash, J. J. Hamlin, R. E. Baumbach, Lijun Zhang, D. J. Singh, M. B. Maple, and D. N. Basov, Nature Physics \textbf{5}, 647 (2009).

\bibitem{LaForge-PRB81-125120-2010} A. D. LaForge, A. Frenzel, B. C. Pursley, Tao Lin, Xinfei Liu, Jing Shi, and D. N. Basov, Phys. Rev. B \textbf{81}, 125120 (2010).

\bibitem{Shushkov-PRB82-125110-2010} A. B. Sushkov, G. S. Jenkins, D. C. Schmadel, N. P. Butch, J. Paglione, and H. D. Drew, Phys. Rev. B. \textbf{82}, 125110 (2010).

%

\bibitem{taskin} A. A. Taskin and Yoichi Ando, Phys. Rev. B \textbf{80}, 085303 (2009).

\bibitem{Zhang-PRB80-085307-2009} Hai-Jun Zhang, Chao-Xing Liu, Xiao-Liang Qi, Xiao-Yu Deng, Xi Dai, Shou-Cheng Zhang, and Zhong Fang, Phys. Rev. B \textbf{80}, 085307 (2009).



\bibitem{Wolff-JPhysChemSolids25-1057-1964} P. A. Wolff, J. Phys. Chem. Solids \textbf{25}, 1057 (1964).

\bibitem{Brown-PRL5-243-1960} R. N. Brown, J. G. Mavroides, M. S. Dresselhaus, and B. Lax, Phys. Rev. Lett. \textbf{5}, 243 (1960).

\bibitem{Vecchi-PRB14-298-1976} M. P. Vecchi, J. R. Pereira, and M. S. Dresselhaus, Phys. Rev. B \textbf{14}, 298 (1976).

\bibitem{Furdyna-ProgPhys33-1193-1970} E. D. Palik and J. K. Furdyna, Rep. Prog. Phys. \textbf{33}, 1193 (1970).

\bibitem{Liu-PRB82-045122-2010} Chao-Xing Liu, Xiao-Liang Qi, Hai-Jun Zhang, Xi Dai, Zhong Fang, and Shou-Cheng Zhang, Phys. Rev. B 82, 045122 (2010)


\bibitem{Fuseya-JPSJ81-013704-2012} Y. Fuseya, M. Ogata, and H. Fukuyama, J. Phys. Soc. of Japan \textbf{81}, 013704 (2012).









\end{thebibliography}
\end{document}